\documentclass[aps,pra,reprint,showpacs]{revtex4-1}
\pdfoutput=1

\usepackage[pdftex]{graphicx}
\usepackage{amsmath}
\usepackage{amsfonts}
\usepackage{mathrsfs}
\usepackage{bm}
\usepackage{hyperref}

\begin{document}

\title{Induced interactions in the BCS-BEC crossover of two-dimensional Fermi gases with Rashba spin-orbit coupling}

\author{Juhee Lee}
\affiliation{Department of Physics and Photon Science, School of Physics and Chemistry, Gwangju Institute of Science and Technology, Gwangju 61005, Korea}
\author{Dong-Hee Kim}
\email{dongheekim@gist.ac.kr}
\affiliation{Department of Physics and Photon Science, School of Physics and Chemistry, Gwangju Institute of Science and Technology, Gwangju 61005, Korea}

\begin{abstract}
We investigate the Gorkov--Melik-Barkhudarov (GM) correction to superfluid transition temperature in two-dimensional Fermi gases with Rashba spin-orbit coupling (SOC) across the SOC-driven BCS-BEC crossover. In the calculation of the induced interaction, we find that the spin-component mixing due to SOC can induce both of the conventional screening and additional antiscreening contributions that interplay significantly in the strong SOC regime. While the GM correction generally lowers the estimate of transition temperature, it turns out that at a fixed weak interaction, the correction effect exhibits a crossover behavior where the ratio between the estimates without and with the correction first decreases with SOC and then becomes insensitive to SOC when it goes into the strong SOC regime. We demonstrate the applicability of the GM correction by comparing the  zero-temperature condensate fraction with the recent quantum Monte Carlo results.
\end{abstract}

\maketitle

\section{Introduction}

The synthetic spin-orbit coupling (SOC) in ultracold gas systems has attracted a lot of attention because of its importance in realizing a new tunable platform for nontrivial condensed matter phenomena~\cite{Dalibard2011,Galitski2013,Dalibard2015,Zhang2014,Yi2015,Zhai2015}. 
For instance, the presence of SOC is an essential part of topological insulators and superconductors~\cite{Hasan2010,Qi2011} and systems for quantum anomalous Hall effect~\cite{Nagaosa2010} and topological quantum computation~\cite{Nayak2008}. The realization of SOC in ultracold atomic gases has rapidly progressed in recent years. The equal mixing of Rashba and Dresselhaus SOC has been realized firstly with bosons~\cite{Lin2011} and 
then with fermions~\cite{Cheuk2012,Wang2012,Williams2013}. Many proposals have been suggested for experimental realization of Rashba-only SOC~\cite{Ruseckas2005,Juzeliunas2010,Campbell2011,Campbell2016,Anderson2013,Xu2013,Liu2014}. Very recently, the realization of two-dimensional SOC which can be transformed into either Rashba or Dresselhaus SOC has been reported~\cite{Huang2016,Meng2016,Wu2016}. 

In attractively interacting Fermi gases, spin-orbit coupling can largely affect the formation of fermion pairs in transition to superfluid (for reviews, see, e.g.,~\cite{Zhang2014,Yi2015,Zhai2015}). In particular, tuning the strength of Rashba SOC can produce another type of crossover between a Bardeen-Cooper-Schrieffer (BCS) superfluid and a Bose-Einstein condensate (BEC) of tightly bound molecules. In the Fermi gases without SOC, the BCS-BEC crossover is typically implemented by controlling interparticle scattering length throughout the Feshbach resonance~\cite{Bartenstein2004,Bourdel2004,Chin2004,Partridge2005,Zwierlein2005}. In the presence of the SOC, it has been suggested that a BEC can be realized even at a fixed weak interaction where very strong SOC can lead to the formation of the bosonic bound state called rashbon~\cite{Vyasanakere2011a,Vyasanakere2011b,Vyasanakere2012}. 
The BCS-BEC crossover in spin-orbit-coupled Fermi gases has been studied with various settings of SOC, interaction, and dimensionality in the mean-field theory~\cite{Hu2011,Gong2011,Yu2011,Iskin2011,Yi2011,Zhou2011,Dellanna2011,Chen2012,He2012b,Han2012,YiXiang2014},  also in the beyond-mean-field approaches for three~\cite{He2013,Vyasanakere2015,Wu2015,Anderson2015,Wang2015} and two dimensions~\cite{He2012a,Zhou2012,Gong2012,Devreese2014}, and very recently by using the quantum Monte Carlo (QMC) method for the ground state in two dimensions~\cite{Shi2016}.

In this paper, we investigate a many-body correction to the estimates of superfluid transition temperature in two-dimensional Fermi gases with Rashba SOC by employing the Gorkov--Melik-Barkhudarov (GM) approach~\cite{GM,Heiselberg2000}. The GM correction incorporates the induced interaction due to the second-order particle-hole processes in the Fermi sea, providing a simple extension of the BCS mean-field framework. In two-component Fermi gases with attractive $s$-wave interaction, without SOC, the screening effect of the induced interaction reduces critical temperature, for instance, by a ratio of about $2.22$ in three dimensions~\cite{GM,Heiselberg2000} and about $2.72$ in two dimensions~\cite{Petrov2003,Resende2012}. 
The GM correction has been also extended to systems with mass imbalance~\cite{Baranov2008}, optical lattices~\cite{Kim2009}, BEC-BCS crossover~\cite{Yu2009}, spin-density polarization~\cite{Yu2010}, three-component gases~\cite{Martikainen2009}, polar molecules~\cite{Baranov2002,Levinsen2011,Fedorov2016}, and spin-orbit coupling~\cite{Caldas2013,Lee2016}.

Here we address two important properties of SOC that make the calculations of the induced interaction essentially different from the conventional case without SOC.   
First, in the presence of SOC, the spin components are mixed, forming the two helicity branches of energy dispersion. These mixed spin states can mediate an antiscreening particle-hole polarization in addition to the conventional screening one, which is analogous to the three-component gases where the third component induces such effect~\cite{Martikainen2009}. Second, the mixed-spin-state character of the Fermi sea abruptly changes across the SOC-driven BCS-BEC crossover~\cite{Vyasanakere2011b}, which may substantially affect the medium contributions to the induced interaction. However, these properties are not included in the earlier consideration of the induced interaction for a spin-orbit-coupled system~\cite{Caldas2013}. 

In the calculation of superfluid transition temperature in the weakly interacting limit, we find that the correction effect with the induced interaction exhibits a crossover behavior between the weak and strong SOC regimes which is attributed to the change of the Fermi sea character. At zero SOC, the transition temperature is confirmed to be reduced with the correction by a ratio of about $2.7$. As the SOC strength increases, this correction ratio decreases in the weak SOC regime but then becomes insensitive to SOC in the strong SOC regime. The ratio in the strong SOC limit is estimated to be about $1.5$ for the mean-field transition temperature, however the correction effect turns out to be much more suppressed in the estimate of the Berezinskii-Kosterlitz-Thouless transition temperature. In addition, we calculate the condensate fraction for comparison with the recent ground-state QMC results~\cite{Shi2016}, showing significant improvement with the GM correction to the mean-field estimate.

This paper is organized as follows. In Sec.~\ref{sec:helicity}, we describe the weak and strong SOC regimes in association with the change in the noninteracting Fermi sea character in the two-dimensional Fermi gas with Rashba SOC. Also in Sec.~\ref{sec:helicity}, the mean-field approximation to estimate the superfluid transition temperature is briefly reviewed. In Sec.~\ref{sec:GM}, we provide the detailed procedures of our induced interaction calculation and the correction to the mean-field equations. In Sec.~\ref{sec:results}, we present the correction to the estimates of superfluid transition temperature in the weakly interacting limit as a function of the SOC strength. The comparison with the QMC results for the condensate fraction is given in this section. In Sec.~\ref{sec:conclusions}, summary and conclusions are given.

\section{Rashba SOC and Helicity branch}
\label{sec:helicity}

\subsection{Helicity basis transformation}

The Hamiltonian of the Rashba spin-orbit-coupled Fermi gases in two dimensions can be written as
\begin{multline}
\mathcal{H} = \sum_{\mathbf{k},\sigma=\uparrow,\downarrow} \xi_k c^\dagger_{\mathbf{k},\sigma} c_{\mathbf{k},\sigma}
- \sum_{\mathbf{k}} \left[ h_R(\mathbf{k}) c^\dagger_{\mathbf{k},\downarrow} c_{\mathbf{k},\uparrow} + \mathrm{h.c.} \right] \\
+ g \sum_{\mathbf{k},\mathbf{k}^\prime} c^\dagger_{\mathbf{k},\uparrow} c^\dagger_{-\mathbf{k},\downarrow} c_{-\mathbf{k}^\prime,\downarrow} c_{\mathbf{k}^\prime,\uparrow}
\end{multline} 
where $c^\dagger_{\mathbf{k},\sigma}$ ($c_{\mathbf{k},\sigma}$) is the creation (annihilation) operator of a $\sigma$-species fermion with momentum $\mathbf{k}\equiv (k_x,k_y)$, and the dispersion $\xi_k = \frac{\hbar^2 k^2}{2m} - \mu$ for chemical potential $\mu$. The volume is set to be unity for simplicity. We consider an attractive $s$-wave interaction between the $\uparrow$- and $\downarrow$-spin species, and thus a negative value is given to the interaction strength $g$. The Rashba SOC is given by the spin-off-diagonal term with $h_R(\mathbf{k}) \equiv v_R ( -k_y + i k_x)$ where $v_R$ indicates the SOC strength. 

The noninteracting part of the Hamiltonian is diagonalized in the $(\Uparrow,\Downarrow)$-helicity basis defined by the unitary transformation written as
\begin{equation}
\begin{pmatrix}
c_{\mathbf{k},\uparrow} \\
c_{\mathbf{k},\downarrow}
\end{pmatrix} 
= \frac{1}{\sqrt{2}}
\begin{pmatrix}
1 & -e^{-i\varphi_\mathbf{k}} \\
e^{i\varphi_\mathbf{k}} & 1 
\end{pmatrix}
\begin{pmatrix}
a_{\mathbf{k},\Uparrow} \\
a_{\mathbf{k},\Downarrow}
\end{pmatrix},
\end{equation}
where $e^{i\varphi_\mathbf{k}} = h_R(\mathbf{k})/|h_R(\mathbf{k})|$. The resulting noninteracting energy dispersions with SOC are obtained as
\begin{eqnarray}
\xi_{\mathbf{k},\Uparrow} &=& \xi_k - v_R |\mathbf{k}|, \\
\xi_{\mathbf{k},\Downarrow} &=& \xi_k + v_R |\mathbf{k}|,
\end{eqnarray} 
which are referred to as the $\Uparrow$- and $\Downarrow$-helicity branches, respectively. For the simplicity of calculations, we use the natural unit $\hbar=2m=1$, and the Boltzmann constant $k_B$ is set to be unity. The particle density $n$ is fixed at $1/2\pi$ throughout our calculations, which also sets the Fermi momentum in the absence of SOC ($v_R=0$) to be unity as $k_\mathrm{F}=\sqrt{2\pi n}=1$. 
This is equivalent to setting the unit of momentum and energy to be $k_\mathrm{F}$ and $\epsilon_\mathrm{F} \equiv \hbar^2 k_\mathrm{F}^2 / 2m$. The SOC strength $v_R$ is accordingly expressed in the unit of $\epsilon_\mathrm{F}/k_\mathrm{F}$.

\begin{figure}
\includegraphics[width=0.45\textwidth]{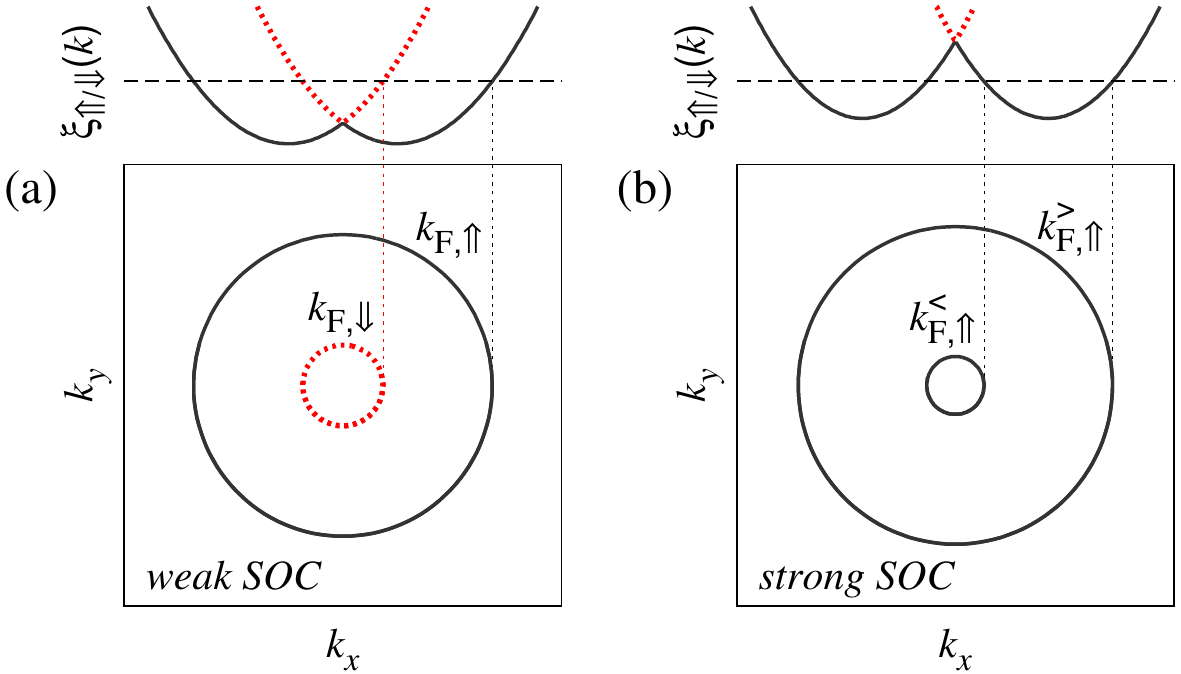}
\caption{Fermi surfaces of two-dimensional Fermi gases with Rashba SOC in noninteracting limit. The solid (dotted) line presents the $\Uparrow$($\Downarrow$) helicity branch of the energy dispersion. The horizontal dashed lines indicate the Fermi energies for the SOC strengths $v_R=1$ (a) and $1.7$ (b) that belong to the weak and strong SOC regimes, respectively.}
\label{fig:FS}
\end{figure}

\subsection{The weak and strong SOC regime}

At a given particle density, the character of the noninteracting Fermi sea abruptly changes as the SOC strength $v_R$ increases from a small to large value as illustrated in Fig.~\ref{fig:FS}. 
At a small $v_R$, both of the $\Uparrow$- and $\Downarrow$-helicity branches can be occupied, and thus the Fermi sea is composed of two different helicity branches. In contrast, at a large $v_R$ the Fermi sea transforms into a doughnutlike shape with two surfaces of the same $\Uparrow$-helicity branch. For $k_\mathrm{F}=\sqrt{2\pi n}=1$, one can easily verify that this transition occurs at $v_R=\sqrt{2}$. For $v_R < \sqrt{2}$, the radii of the $\Uparrow$- and $\Downarrow$-helicity Fermi seas are
\begin{eqnarray}
k_{\mathrm{F},\Uparrow} &=& v_R/2+\sqrt{1-(v_R/2)^2}, \\
k_{\mathrm{F},\Downarrow} &=& -v_R/2+\sqrt{1-(v_R/2)^2}.
\end{eqnarray}
On the other hand, for $v_R > \sqrt{2}$, the Fermi sea is surrounded by the two surfaces with radii $k_{\mathrm{F},\Uparrow}^>$ and $k_{\mathrm{F},\Uparrow}^<$ which are obtained as 
\begin{eqnarray}
k_{\mathrm{F},\Uparrow}^> &=& v_R/2 + 1/v_R, \\
k_{\mathrm{F},\Uparrow}^< &=& v_R/2 - 1/v_R.
\end{eqnarray} 
In order to distinguish these two different characters of the noninteracting Fermi sea, we refer to the SOC strengths of $v_R<\sqrt{2}$ and $v_R>\sqrt{2}$ as the weak and strong SOC regimes, respectively.

\subsection{Mean-field approach to superfluid transition}

The GM correction procedures are based on a perturbative approach that preserves the structure of the mean-field theory. The many-body effect induced by the Fermi sea is considered as an effective interaction. With the superfluid order parameter $\Delta \equiv g\sum_{\mathbf{k}} \langle c_{-\mathbf{k},\downarrow} c_{\mathbf{k},\uparrow} \rangle$, the mean-field Hamiltonian is written as
\begin{equation}
\mathcal{H}_\mathrm{MF} = \frac{1}{2} \sum_{\mathbf{k}} \Psi^\dagger_\mathbf{k} \widetilde{\mathcal{H}}_\mathbf{k} \Psi_\mathbf{k} + \sum_{\mathbf{k}} \xi_k - \frac{|\Delta|^2}{g},
\end{equation}
where $\Psi^\dag_\mathbf{k} = (c^\dag_{\mathbf{k},\uparrow},c^\dag_{\mathbf{k},\downarrow},c_{-\mathbf{k},\uparrow},c_{-\mathbf{k},\downarrow})$, and the matrix 
\begin{equation}
\widetilde{\mathcal{H}}_\mathbf{k} =
\begin{pmatrix}
\xi_k & -h_R^*(\mathbf{k}) & 0 & \Delta \\
-h_R(\mathbf{k}) & \xi_k & -\Delta & 0 \\
0 & -\Delta^* & -\xi_k & -h_R(\mathbf{k}) \\
\Delta^* & 0 & -h_R^*(\mathbf{k}) & -\xi_k
\end{pmatrix}.
\end{equation}
Diagonalizing $\widetilde{\mathcal{H}}(\mathbf{k})$, one can obtain the quasiparticle energies as $E_{k,\pm} = \sqrt{(\xi_k \pm v_R k)^2+|\Delta|^2}$, and accordingly one can write down the thermodynamic potential as 
\begin{equation*}
\label{eq:potential}
\Omega = \sum_{\mathbf{k},s=\pm} \left[ \frac{\xi_k}{2} - \frac{E_{k,s}}{2}  
- \frac{1}{\beta} \ln \left( 1 + e^{-\beta E_{k,s}} \right) \right]
- \frac{|\Delta|^2}{g}.
\end{equation*}
From the saddle-point condition $\frac{1}{\Delta}\frac{\partial \Omega}{\partial \Delta^*}=0$, the mean-field gap equation is derived as
\begin{equation}
\label{eq:gap_BCS}
\int \frac{d^2 k}{(2\pi)^2} \left[ \frac{1}{2\epsilon_k + \epsilon_B} - \sum_{s=\pm} \frac{\tanh \frac{\beta E_{k,s}}{2}}{4E_{k,s}} \right] = 0,
\end{equation}
where $\epsilon_k=\hbar^2k^2/2m$. Note that the bare interaction strength $g$ is replaced by the two-body binding energy $\epsilon_B$ through the relation for a two-dimensional gas
\begin{equation}
-\frac{1}{g} = \int \frac{d^2 k}{(2\pi)^2} \frac{1}{2\epsilon_k + \epsilon_B}.
\end{equation}
The number equation $n=-\frac{\partial \Omega}{\partial \mu}$ is also written as
\begin{equation}
\label{eq:number}
n = \int \frac{d^2 k}{(2\pi)^2} \left[ 1-\sum_{s=\pm} \left( \xi_k + s v_R k \right) \frac{\tanh \frac{\beta E_{k,s}}{2}}{2E_{k,s}} \right].
\end{equation}
For a given particle density, the superfluid transition temperature is determined by self-consistently solving the gap and number equations for the vanishing order parameter. The modification by the GM correction is to be brought only into the mean-field gap equation by replacing the bare interaction $g$ with the effective interaction $\bar{g}_\mathrm{eff}$ that includes the induced interaction correction. 

\section{Induced Interactions}
\label{sec:GM}

\begin{figure}
\includegraphics[width=0.48\textwidth]{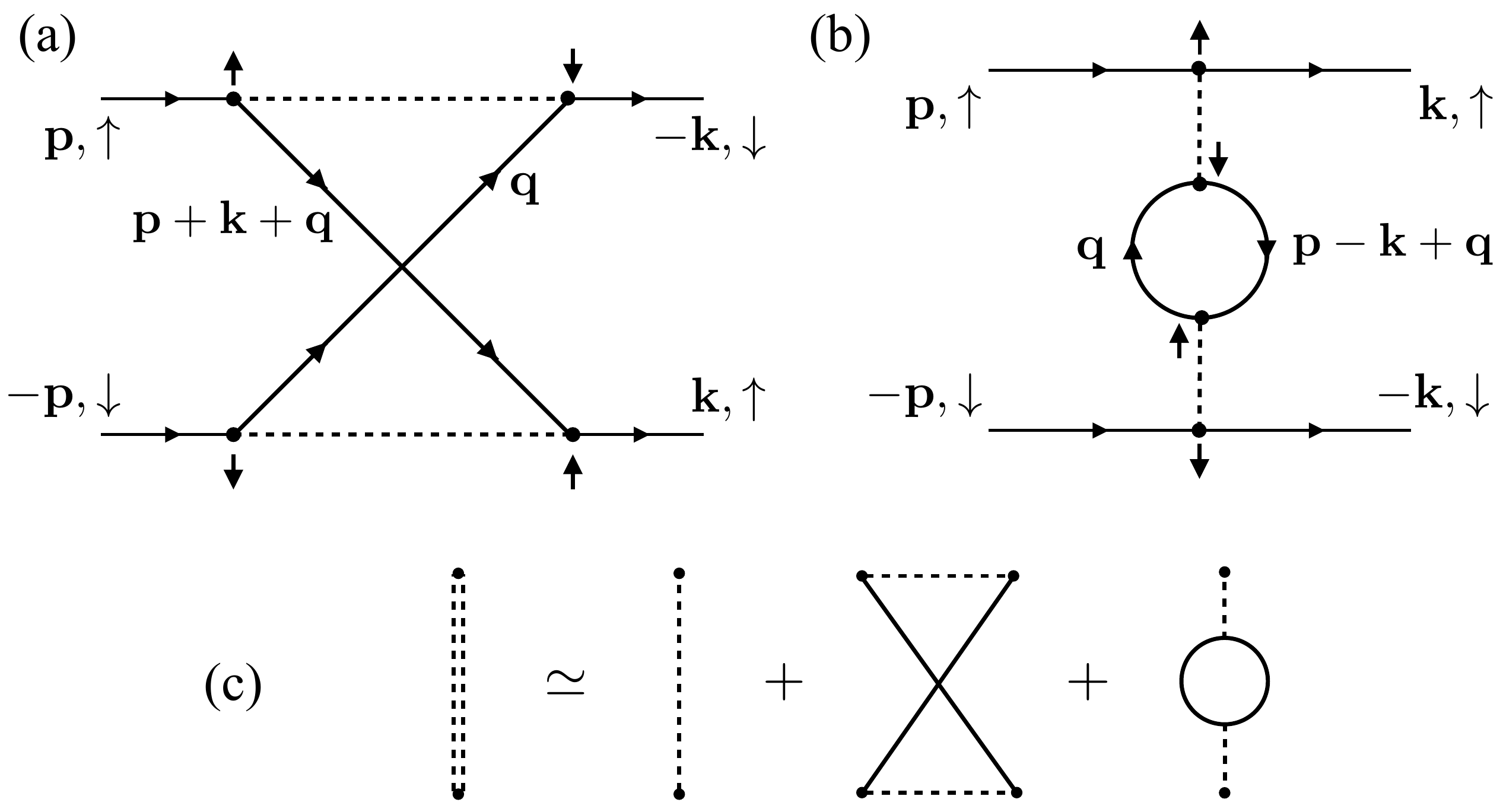}
\caption{Second-order diagrams of the medium-induced interaction between the $\uparrow$- and $\downarrow$-spin components. The solid lines are propagators, and the single dotted line indicates the bare two-body interaction. The diagram in (a) is conventional in usual two-component systems, and the one in (b) is mediated by the spin-component mixing due to SOC. (c) The schematic diagram of the effective interaction (double dotted line) with the induced interaction correction.}
\label{fig:diagrams}
\end{figure}

Figure~\ref{fig:diagrams} illustrates the two relevant second-order diagrams of the induced interaction in the presence of SOC. While diagram $1$ shown in Fig.~\ref{fig:diagrams}(a) is typical in a two-component system with an $s$-wave interaction, diagram $2$ shown in Fig.~\ref{fig:diagrams}(b) is allowed by the spin-offdiagonal propagators available in the mixed spin states due to the SOC. The particle-hole polarizations presented in diagrams $1$ and $2$ contribute to the induced interaction correction. In the GM correction scheme, the effective interaction $\bar{g}_{\mathrm{eff}}$ is written with the correction up to the second-order of interaction  as  
\begin{equation}
\bar{g}_{\mathrm{eff}} = g + \bar{g}_{\mathrm{ind}} \simeq g - g^2 \left( \overline{\Pi}_1 + \overline{\Pi}_2 \right),
\end{equation} 
where the induced interaction $\bar{g}_{\mathrm{ind}}\equiv- g^2 \left( \overline{\Pi}_1 + \overline{\Pi}_2 \right)$ is to be averaged over the Fermi surfaces that are expected to dominantly contribute to the fermion pairing.

The contributing polarization functions $\Pi_1$ and $\Pi_2$ of diagrams 1 and 2 are expressed by using propagators in an imaginary-frequency domain as 
\begin{eqnarray}
\Pi_1 (\tilde{p},\tilde{k}) &=& \frac{1}{\beta} \sum_{\tilde{q}} \mathcal{G}^{(0)}_{\uparrow \uparrow} (\tilde{p}+\tilde{k}+\tilde{q}) \mathcal{G}^{(0)}_{\downarrow \downarrow} (\tilde{q}), \\
\Pi_2 (\tilde{p},\tilde{k}) &=& -\frac{1}{\beta} \sum_{\tilde{q}} \mathcal{G}^{(0)}_{\uparrow \downarrow} (\tilde{p}-\tilde{k}+\tilde{q}) \mathcal{G}^{(0)}_{\downarrow \uparrow} (\tilde{q}),
\end{eqnarray}
where the four-momentum $\tilde{z} \equiv (\mathbf{z},iz_n)$ is fermionic, and the noninteracting Matsubara Green's function $\mathcal{G}^{(0)}_{\sigma\sigma'}(\tilde{z})=-\int_0^\beta \left\langle T_{\tau} \left( c_{\mathbf{z}\sigma}(\tau)c^\dagger_{\mathbf{z}\sigma'}(0) \right) \right\rangle_0 e^{iz_n \tau}d\tau$.
The spin-diagonal and offdiagonal parts of $\mathcal{G}^{(0)}_{\sigma\sigma^\prime}$ can be evaluated through the helicity basis transformation as
\begin{eqnarray}
\mathcal{G}^{(0)}_{\sigma\sigma} (\tilde{k}) &=& \frac{1}{2} \left[ \mathscr{G}_{0\Uparrow} (\tilde{k}) + \mathscr{G}_{0\Downarrow} (\tilde{k}) \right], \\
\mathcal{G}^{(0)}_{\sigma\bar{\sigma}} (\tilde{k}) &=& \frac{1}{2} e^{s_{\sigma\bar{\sigma}} i\varphi_\mathbf{k}} \left[ \mathscr{G}_{0\Uparrow} (\tilde{k}) - \mathscr{G}_{0\Downarrow} (\tilde{k}) \right],
\end{eqnarray}
where $s_{\uparrow\downarrow} = -s_{\downarrow\uparrow} = 1$, and $\mathscr{G}_{0a} (\tilde{k}) = 1/(ik_n -\xi_{\mathbf{k},a})$ for $a\in \{\Uparrow,\Downarrow\}$.
Then, $\Pi_1$ and $\Pi_2$ can be decomposed with the inter- and intra-helicity-branch components as 
\begin{equation}
\Pi_{1,2}(\mathbf{p},\mathbf{k}) = \sum_{a,b \in \{\Uparrow,\Downarrow\}} \left[ \Pi_{1,2}(\mathbf{p},\mathbf{k}) \right]_{ab}
\label{eq:PI}
\end{equation}
where
\begin{eqnarray}
\left[ \Pi_1 (\mathbf{p},\mathbf{k}) \right]_{ab} &=& \frac{1}{4} \int\frac{d^2 q}{(2\pi)^2} \chi_{ab} (\mathbf{p}+\mathbf{k},\mathbf{q}), \label{eq:PI1} \\
\left[ \Pi_2 (\mathbf{p},\mathbf{k}) \right]_{ab} &=& -\frac{c_{ab}}{4} \int\frac{d^2 q}{(2\pi)^2} e^{i\Delta\varphi}  \chi_{ab} (\mathbf{p}-\mathbf{k},\mathbf{q}) \label{eq:PI2}.
\end{eqnarray}
The coefficient $c_{ab}$ is $+1$ for $a=b$ or $-1$ for $a \ne b$, and $\Delta\varphi = \varphi_{\mathbf{q}}-\varphi_{\mathbf{p}-\mathbf{k}+\mathbf{q}}$. The function $\chi_{ab}$ is calculated by performing imaginary-frequency summation as
\begin{eqnarray}
\chi_{ab} (\mathbf{p}\pm\mathbf{k},\mathbf{q}) &=& \frac{1}{\beta}\sum_{iq_n} \mathscr{G}_{0a} (\tilde{p}\pm\tilde{k}+\tilde{q}) \mathscr{G}_{0b} (\tilde{q}) \\
&=& \frac{n_F(\xi_{\mathbf{p}\pm\mathbf{k}+\mathbf{q},a})-n_F(\xi_{\mathbf{q},b})}{i\Omega + \xi_{\mathbf{p}\pm\mathbf{k}+\mathbf{q},a} - \xi_{\mathbf{q},b}} \nonumber
\label{eq:chi}
\end{eqnarray}
where $n_F(\xi) = 1/(e^{\beta\xi}+1)$, and the bosonic frequency $\Omega$ is taken to be zero in the low temperature limit in the evaluation of the effective interaction. Our formulation of the induced interaction extends the previous approach~\cite{Caldas2013} where only the interbranch  $\chi_{\Uparrow\Downarrow}$ was considered within diagram $1$. Later, we will show that the contribution of $\chi_{\Uparrow\Downarrow}$ is rather small at strong SOC.

In the standard procedures of the GM correction, the induced interaction is evaluated as $\Pi$ being averaged over the Fermi surface momenta~\cite{Heiselberg2000}, which can be justified in the weakly interacting limit where the scattering processes for pairing would dominantly occur near the Fermi surface. Indeed, the recent QMC calculations for two-dimensional Fermi gases with Rashba SOC showed that the condensate wave functions are peaked at the two Fermi surfaces in the weak and strong SOC regimes~\cite{Shi2016}. Therefore, we evaluate the induced interaction as 
\begin{eqnarray}
\bar{g}_{\mathrm{ind}} &\equiv & -g^2 \left[ \overline{\Pi}_1 + \overline{\Pi}_2 \right] \nonumber \\
&=& -g^2 \left[ \left\langle \Pi_1 (\mathbf{p},\mathbf{k}) \right\rangle_{\mathrm{FS}} + \left\langle \Pi_2 (\mathbf{p},\mathbf{k}) \right\rangle_{\mathrm{FS}} \right],
\end{eqnarray}
where $\langle \Pi \rangle_\mathrm{FS} = \frac{1}{||\mathcal{S}_F||^2}\sum_{\mathbf{p}\in \mathcal{S}_F} \sum_{\mathbf{k} \in \mathcal{S}_F} \Pi(\mathbf{p},\mathbf{k})$, and $\mathcal{S}_F$ denotes a set of the Fermi surface momenta. We treat the two Fermi surfaces on an equal footing for averaging, and thus $\mathcal{S}_F$ includes all momenta residing on the two surfaces composed of the one with $k_{\mathrm{F},\Uparrow}$ ($k_{\mathrm{F},\Uparrow}^{>}$) and the other with $k_{\mathrm{F},\Downarrow}$ ($k_{\mathrm{F},\Uparrow}^{<}$) in the weak (strong) SOC regime. All momentum integrations are done numerically for the evaluation of the induced interaction.

The GM correction to the mean-field gap equation is then readily done by replacing bare interaction $g$ with effective interaction $\bar{g}_{\mathrm{eff}}$. 
In the weakly interacting limit, the inverse of the effective interaction becomes 
\begin{equation}
\frac{1}{\bar{g}_{\mathrm{eff}}} = \frac{1}{g - g^2 \left( \overline{\Pi}_1 + \overline{\Pi}_2 \right)} \simeq \frac{1}{g} + \overline{\Pi}_1 + \overline{\Pi}_2,
\label{eq:g_eff}
\end{equation}
which rewrites the mean-field gap equation~(\ref{eq:gap_BCS}) as
\begin{equation}
\int \frac{d^2 k}{(2\pi)^2} \left[ \frac{1}{2\epsilon_k + \epsilon_B} - \sum_{s=\pm} \frac{\tanh \frac{\beta E_{k,s}}{2}}{4E_{k,s}} \right] = \overline{\Pi}_1 + \overline{\Pi}_2.
\label{eq:gap_GM}
\end{equation} 
The applicability of the GM-corrected gap equation mainly depends on the interaction strength since the induced interaction is basically a second-order perturbation expansion for a weak interaction. In addition, we have assumed that all intra- and inter-surface pairs of the Fermi momenta equally participate in the Fermi surface average for the induced interaction. These approximations may need to be examined in more rigorous approaches. We will examine the improvement with the correction by comparing with the recent ground-state QMC results for the condensate fraction~\cite{Shi2016} in the following section.

\section{Results and Discussions}
\label{sec:results}

\subsection{Induced interactions}

\begin{figure}
\includegraphics[width=0.48\textwidth]{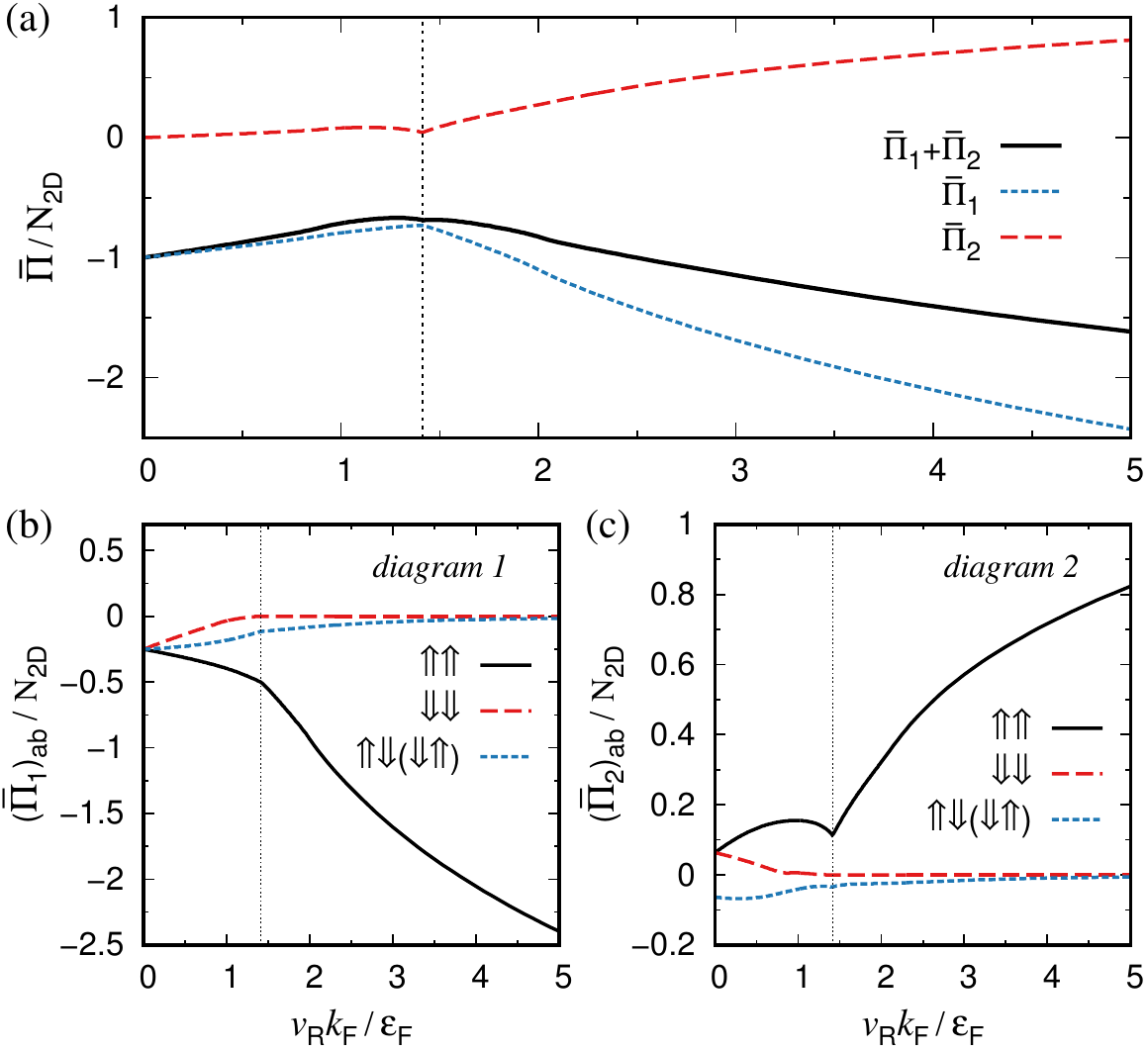}
\caption{Induced interaction as a function of the SOC strength. The values are scaled with the density of states $\mathcal{N}_{\mathrm{2D}}$ of two-dimensional Fermi gases without SOC. (a) Total induced interaction $\bar{g}_{\mathrm{ind}}/(-g^2)\equiv \overline{\Pi}_1+\overline{\Pi}_2$, and the screening and antiscreening contributions $\overline{\Pi}_1$  and $\overline{\Pi}_2$ from diagrams $1$  and $2$ given in Fig.~\ref{fig:diagrams}, respectively. The inter- and intra-helicity-branch components $(\overline{\Pi})_{ab}$ are presented for diagrams $1$ (b) and $2$ (c). The vertical dotted lines indicate $v_R=\sqrt{2}$ where the character of the Fermi sea abruptly changes.}
\label{fig:PI}
\end{figure}

Figure~\ref{fig:PI} presents the induced interaction calculated from diagrams $1$ and $2$ as a function of the SOC strength $v_R$. 
The total induced interaction $\bar{g}_{\mathrm{ind}}=-g^2(\overline{\Pi}_1+\overline{\Pi}_2)$ is positive for all $v_R$ and thus reduces the bare attraction $g$ in the effective interaction $\bar{g}_\mathrm{eff} \equiv g + \bar{g}_\mathrm{ind}$. This would lower the estimation of superfluid transition temperature. However, we find that the induced interaction and its components exhibit a nontrivial SOC-strength dependence. This appears as a crossover from the weak to strong SOC regime that is attributed to the abrupt change in the mixed-spin-state character of the noninteracting Fermi sea.

In the weak SOC regime, the induced interaction is mainly determined by the screening effect of $\overline{\Pi}_1$ from the contribution of diagram $1$. In the limit of $v_R \to 0$, all components of the polarization function, $\chi_{ab}$ given in Eq.~(\ref{eq:chi}), become identical, leading to the complete cancellation in $\Pi_2$, while in turn $\Pi_1$ recovers the case without SOC. The negative sign of $\overline{\Pi}_1$ can be easily understood from the form of $\chi_{ab}$ which is similar to the conventional two-component systems without SOC. As $v_R$ increases, $\overline{\Pi}_2$ becomes finite with the opposite sign indicating an antiscreening effect, although it remains much smaller in magnitude than $\overline{\Pi}_1$ for small $v_R$.  

On the other hand, in the strong SOC regime, the antiscreening contribution of $\overline{\Pi}_2$ is no longer small but considerably weakens the screening effect of $\overline{\Pi}_1$ in the total induced interaction. At $v_R  > \sqrt{2}$, the noninteracting Fermi sea is available only with the $\Uparrow$-helicity branch for particles to reside, which leads to the dominant contribution of the $\chi_{\Uparrow\Uparrow}$ component in both $\Pi_1$ and $\Pi_2$ while the other components are largely suppressed. A kink was found at $v_R = \sqrt{2}$ where the $\Downarrow$-helicity branch disappears in the Fermi sea. The induced interaction is determined by a net effect of the opposite contributions of $\Pi_1$ and $\Pi_2$, and their proper consideration becomes very important for the evaluation of the GM correction in spin-orbit-coupled systems. 

\subsection{Correction to superfluid transition temperature}

\begin{figure}
\includegraphics[width=0.48\textwidth]{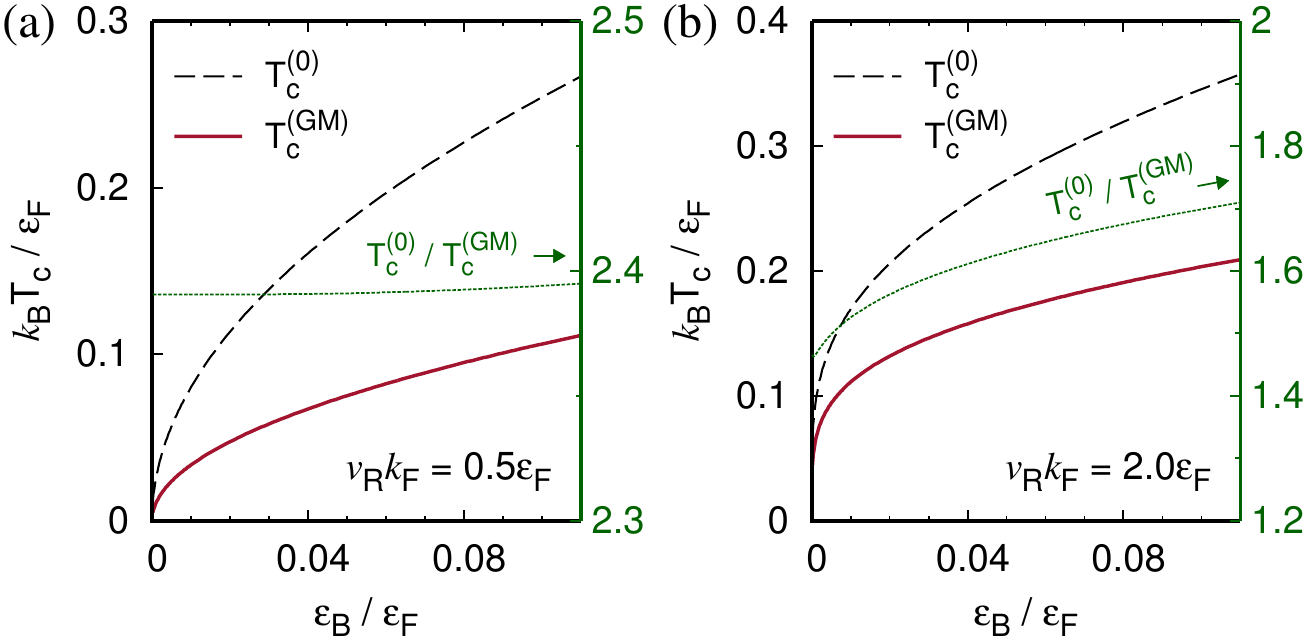}
\caption{Induced-interaction correction to the mean-field transition temperature as a function of the interaction strength. The ratio of the transition temperature without the correction ($T_c^{(0)}$) and with the correction ($T_c^{(\mathrm{GM})}$) is compared between the weak and strong SOC regimes at (a) $v_R =0.5$ and (b) $v_R =2.0$, respectively.}
\label{fig:tc}
\end{figure}

We calculate the correction to the estimates of superfluid transition temperature in the weakly interacting limit. First, we present the effects of the induced interaction correction on the mean-field calculations which can be more relevant in a quasi-two-dimensional system supporting the long-range order.  Additionally, for a strictly two-dimensional system where the long-range order is absent, we provide a rough estimate of the Berezinskii-Kosterlitz-Thouless transition temperature under the assumption that the induced interaction effect is limited to the mean-field gap amplitude. The influence of the induced interaction correction varies with the strength of SOC, which we discuss below in terms of the ratio between the estimates of the transition temperature without and with the correction. 

We first examine the interaction dependence of the GM correction to the superfluid transition temperature in the weak and strong SOC regimes. Figure~\ref{fig:tc} shows the comparison between the mean-field transition temperature estimates $T_c^{(0)}$ without the correction and $T_c^{(\mathrm{GM})}$ with the correction and the reduction ratio $T_c^{(0)}/T_c^{(\mathrm{GM})}$ as a function of $\epsilon_B$. In both of the values of $v_R$ that we have examined, the reduction ratio goes below the known zero-SOC value of about $2.7$ in the weakly interacting limit and also shows the weaker influence of the correction at the stronger SOC.

\begin{figure*}
\includegraphics[width=0.95\textwidth]{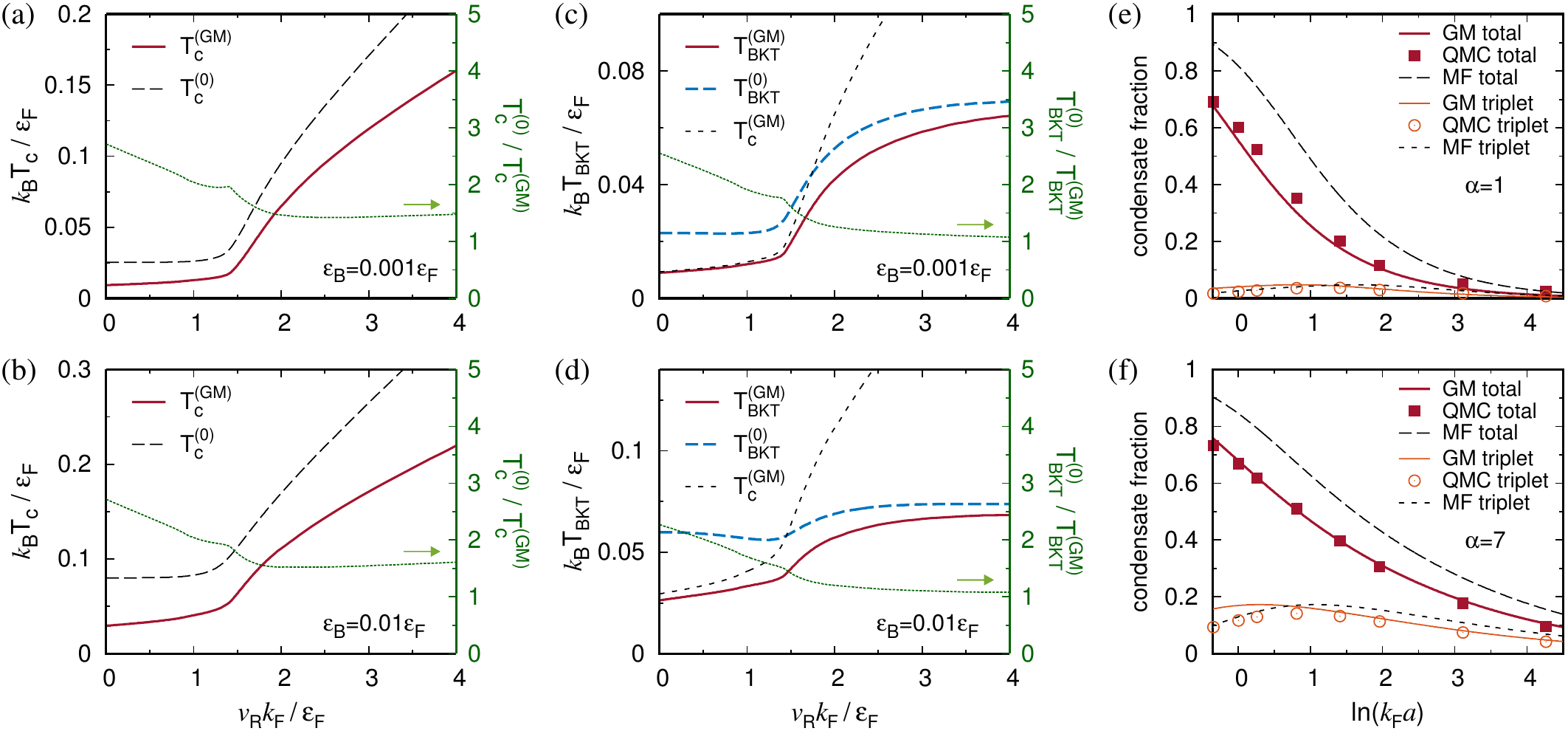}
\caption{Superfluid transition temperature with the GM correction and comparison with the QMC data of condensate fraction. The mean-field transition temperature $T_c$ (a) and (b), and the BKT transition temperature $T_{\mathrm{BKT}}$ (c) and (d), are calculated as a function of the SOC strength $v_R$ at weak attractions of $\epsilon_B=0.001$ and $\epsilon_B=0.01$. The superscripts $(\mathrm{GM})$ and $(0)$ indicate ones with and without the GM correction, respectively. The condensate fraction at zero temperature is compared with the recent QMC results~\cite{Shi2016} for the weak and strong SOC regimes at (e) $\alpha \equiv 2v_R^2/\epsilon_F = 1$ and (f) $\alpha=7$. For direct comparison, the binding energy $\epsilon_B$ is converted into the scattering length $a$ through the relation $\epsilon_B \equiv 4\hbar^2/ma^2e^{2\gamma}$ where $\gamma$ is the Euler's constant. The lines marked by $\mathrm{MF}$ indicate the mean-field values without the GM correction. 
}
\label{fig:main}
\end{figure*}

The variation of the correction effect with the SOC strength is more systematically shown in Figs.~\ref{fig:main}(a) and \ref{fig:main}(b) at the fixed values of $\epsilon_B=0.001$ and $\epsilon_B=0.01$ that are close to the weakly interacting limit. We find that the reduction ratio $T_c^{(0)}/T_c^{(\mathrm{GM})}$ first decreases monotonically from the zero-SOC value of about $2.7$ as $v_R$ increases within the weak SOC regime, but then the ratio becomes insensitive to SOC when $v_R$ goes into the strong SOC regime. A kink appears at $v_R=\sqrt{2}$ where the noninteracting Fermi sea abruptly changes.
While the effect of the induced interaction decreases with increasing SOC in terms of the reduction ratio, it turns out that the ratio at a large $v_R$ approaches a constant value about $1.5$ which still gives a sizable reduction to $T_c^{(0)}$. 

On the other hand, in strictly two dimensions, the true long-ranger order is absent because of strong quantum fluctuations. Although, it is well known that the superfluid transition at finite temperature is still possible by the Berezinskii-Kosterlitz-Thouless (BKT) mechanism of the vortex-antivortex pairing~\cite{Berezinskii1971,KT1972}. In the spin-orbit-coupled Fermi gases, the BKT transition temperature and superfluid properties have been estimated with consideration of the phase fluctuations in the order parameter beyond the mean-field approximation~\cite{Zhou2012,He2012a,Gong2012,Devreese2014}. 

Regarding the GM correction, the following question naturally arises for the BKT transition temperature. At weak interactions, the previous result~\cite{He2012a} shows that the BKT transition temperature recovers the mean-field estimate when the Rashba SOC strength is small, while in the strong SOC limit, the BKT temperature largely deviates from the mean-field estimate. However, we find that for instance, at a very small $v_R$, the estimate of the mean-field transition temperature with the GM correction is reduced by a factor of about $2.7$, and therefore it happens that the mean-field estimate with the correction actually goes much below the BKT temperature if estimated without such correction.

Here we suggest that in the weakly interacting limit, the GM correction may be applicable to the estimate of the BKT temperature by restricting the influence of the induced interaction within the mean-field or saddle-point gap equation. We follow the description of the phase fluctuations given in Ref.~\cite{He2012a}. In the procedures, when the amplitude fluctuations of the order parameter is neglected, the amplitude is still determined from the same mean-field gap equation. Thus, as a rough approximation in the weakly interacting limit, by ignoring any effects on the anomalous propagators, one may include the correction in the same way as for the mean-field calculations with the GM-corrected gap equation. 

The BKT transition temperature ($T_{\mathrm{BKT}}$) can be determined through the universal Nelson-Kosterlitz (NK) relation~\cite{NK1977} which can be written as $T_{\mathrm{BKT}} = \frac{\pi}{4}\rho_s(T_{\mathrm{BKT}})$ in our natural unit. The superfluid density $\rho_s$ is evaluated as $\rho_s(T) = n - \rho_1(T) - \rho_2(T)$~\cite{He2012a} as a function of temperature $T$, where $\rho_1$ and $\rho_2$ in our unit are given as
\begin{eqnarray}
\rho_1(T) &=& \frac{v_R}{16\pi} \sum_{s=\pm} \int_0^\infty dk \, s \left(\xi_{k,s} + \frac{\Delta^2}{\xi_k}\right)\frac{\tanh[\frac{\beta E_{k,s}}{2}]}{E_{k,s}} ,\nonumber \\
\rho_2(T) &=& \frac{\beta}{8\pi} \sum_{s=\pm} \int_0^\infty k dk \left( k + s \frac{v_R}{2} \right)^2 \mathrm{sech}^{2} [\frac{\beta E_{k,s}}{2}] ,\nonumber
\end{eqnarray}
for $\xi_{k,+/-} \equiv \xi_{k,\Downarrow/\Uparrow}$. With the GM-corrected gap equation employed for $\Delta$, one can find the corresponding correction in the estimate of the BKT transition temperature $T_\mathrm{BKT}^{(\mathrm{GM})}$ by solving the NK relation.

The correction in the BKT transition temperature is shown in Figs.~\ref{fig:main}(c) and \ref{fig:main}(d) for the same choices of $\epsilon_B$. With the correction, it turns out that $T_{\mathrm{BKT}}^{(\mathrm{GM})}$ becomes very close to the mean-field estimate $T_c^{(\mathrm{GM})}$ at small $v_R$, recovering the expectation of Ref.~\cite{He2012a}. In the reduction ratio between $T_{\mathrm{BKT}}^{(0)}$, the one without the correction, and $T_{\mathrm{BKT}}^{(\mathrm{GM})}$, the behavior of its SOC dependence exhibits the similar crossover observed in the mean-field result: The ratio decreases with $v_R$ at first and then becomes insensitive to $v_R$ in the strong SOC regime. Interestingly, the asymptotic value of the ratio approaches unity at a large $v_R$, implying that the effect of the induced interaction on the BKT transition temperature may diminish in the Rashbon limit.

\subsection{Comparison with the QMC results}

Finally, we demonstrate the applicability of the GM-corrected gap equation by the comparison with the recent ground-state QMC calculations~\cite{Shi2016} for the condensate fraction.
In the mean-field approximation~\cite{He2012a,Zhou2012}, the condensed density of fermion pairs $n_0$ is expressed as $n_0 = \sum_\mathbf{k} |\phi_{\uparrow\downarrow}(\mathbf{k})|^2 + |\phi_{\uparrow\uparrow}(\mathbf{k})|^2$ where the spin-singlet pairing field $\phi_{\uparrow\downarrow}(\mathbf{k}) \equiv \langle c_{\mathbf{k},\uparrow} c_{-\mathbf{k},\downarrow}\rangle = -\Delta \sum_{s=\pm} 1/4E_{k,s}$ and the SOC-induced spin-triplet pairing field $\phi_{\uparrow\uparrow}(\mathbf{k}) \equiv \langle c_{\mathbf{k},\uparrow} c_{-\mathbf{k},\uparrow}\rangle = -\Delta e^{-i\varphi_{\mathbf{k}}} \sum_{s=\pm} s/4E_{k,s}$ at $T=0$. The total condensate fraction $n_c\equiv 2n_0/n$ is then written as  
\begin{equation}
n_c = \frac{|\Delta |^2}{4} \int_0^\infty k dk \left( \frac{1}{E^2_{k,+}} + \frac{1}{E^2_{k,-}} \right).
\end{equation}
This only depends on the zero-temperature order parameter which is obtained from Eq.~(\ref{eq:gap_GM}) when the GM correction is included.

In Figs.~\ref{fig:main}(e) and \ref{fig:main}(f), we provide the condensate fraction calculated at the two chosen values of the SOC strength for direct comparison with the QMC data in the weak and strong SOC regimes. The total condensate fractions indicate good agreement, showing a significant improvement with the GM correction to the mean-field estimate. The agreement is better in the stronger SOC even at relatively strong interactions, which is notable since the GM correction is generally expected to be most reliable at weak interactions. In the spin-triplet component, the perturbative nature of the GM correction indeed becomes apparent at strong interactions where the QMC data deviate from the GM-corrected one. Such systematic deviation with increasing interaction is not seen in the total condensate fraction at the stronger SOC, implying that the underestimation of the singlet component compensates the difference.

\section{Summary and Conclusions}
\label{sec:conclusions}

We have investigated the Gorkov--Melik-Barkhudarov correction to superfluid transition temperature in attractively interacting Fermi gases with Rashba SOC in two dimensions. The main differences due to the presence of SOC are summarized as follows. First, the spin-component mixing allows another second-order process that additionally leads to the antiscreening effect that is prohibited in the usual two-component case without SOC. The induced interaction is thus determined by a net effect of the antiscreening and conventional screening contributions. Second, the induced interaction and corresponding correction to the superfluid transition temperature largely depends on the SOC strength, which is attributed to the abrupt change of the Fermi sea character across the SOC-driven BCS-BEC crossover. 

In the estimate of the mean-field transition temperature in the weakly interacting limit, the correction due to the induced interaction shows a crossover behavior across the weak and strong SOC regimes. At strong SOC, it is found that the mean-field temperature is reduced by a ratio of about $1.5$ with the GM correction which is in contrast to the known value of about $2.7$ for the case without SOC. This ratio decreases with increasing SOC in the weak SOC regime, but then it becomes largely insensitive to SOC in the strong SOC regime. The similar crossover is also found in our estimation of the BKT transition temperature at weak interactions. However, our calculation indicates that the correction becomes quantitatively very small in the strong SOC limit in the case of the BKT transition temperature.

The essence of the GM correction procedures is in the modification of the mean-field gap equation with the induced interaction. Our comparison with the recent ground-state QMC result for the condensate fraction~\cite{Shi2016} shows remarkable improvement to the use of the usual mean-field order parameter, demonstrating the importance of the GM correction in a spin-orbit-coupled system. While exact numerics with SOC for the transition temperature have not been available yet, our estimates with the GM correction would motivate applications of other beyond-mean-field approaches, for instance, with the functional renormalizaton group~\cite{Floerchinger2008}, particle-hole fluctuations with self-energy renormalization~\cite{Gubbels2009}, non-Gaussian fluctuation theory~\cite{Vyasanakere2015}, and $t$-matrix approaches~\cite{Wu2015,Anderson2015,Chen1998,Haussmann2007,Bauer2014,Chen2016}, to two-dimensional Fermi gases with SOC. In addition, we expect that it is also possible to extend our formulation of the GM correction to systems in three dimensions or with other types of SOC, which would provide more improved mean-field-based predictions of experimental relevance.

\begin{acknowledgments}
We thank Peter Rosenberg and Shiwei Zhang for providing their quantum Monte Carlo data and Jildou Baarsma for valuable discussions. This work was supported from Basic Science Research Program through the National Research Foundation of Korea funded by the Ministry of Science, ICT \& Future Planning (Grant No. NRF-2014R1A1A1002682).
\end{acknowledgments}

\end{document}